\documentclass[10pt]{article}

\usepackage[superscript,sort]{cite}

\usepackage{ifthen,ifpdf}
\ifpdf
	\usepackage[pdftex]{hyperref}	\hypersetup{colorlinks=true,linkcolor=black,citecolor=black,urlcolor=blue,backref=page,bookmarks=true,breaklinks=true,plainpages=false }
	\usepackage[pdftex]{graphicx}
	\usepackage[usenames,pdftex]{color}
\else
	\usepackage[colorlinks=true,breaklinks=true,plainpages=false]{hyperref}
	\usepackage{graphicx}
	\usepackage[usenames]{color}
\fi

\usepackage{amsthm,amscd,amsxtra,amsfonts,amsmath,amssymb,multirow}
\usepackage{wrapfig}
\usepackage[footnotesize]{caption}
\usepackage[tiny,compact]{titlesec}
\usepackage[textwidth=0.8in,textsize=footnotesize]{todonotes}
\usepackage{algorithm,algorithmic,extarrows}

\begin{document}

\title{A quantitative structure comparison with persistent similarity}

\author{
Kelin Xia$^{1,2}$ \footnote{ Address correspondences  to Kelin Xia. E-mail:xiakelin@ntu.edu.sg}\\
$^1$Division of Mathematical Sciences, School of Physical and Mathematical Sciences, \\
 Nanyang Technological University, Singapore 637371 \\
$^2$School of Biological Sciences, Nanyang Technological University, Singapore 637371
}

\date{\today}
\maketitle

\begin{abstract}
Biomolecular structure comparison not only reveals evolutionary relationships, but also sheds light on biological functional properties. However, traditional definitions of structure or sequence similarity always involve superposition or alignment and are computationally inefficient. In this paper, I propose a new method called persistent similarity, which is based on a newly-invented method in algebraic topology, known as persistent homology. Different from all previous topological methods, persistent homology is able to embed a geometric measurement into topological invariants, thus provides a bridge between geometry and topology. Further, with the proposed persistent Betti function (PBF),
topological information derived from the persistent homology analysis can be uniquely represented by a series of continuous one-dimensional (1D) functions. In this way, any complicated biomolecular structure can be reduced to several simple 1D PBFs for comparison. Persistent similarity is then defined as the quotient of sizes of intersect areas and union areas between two correspondingly PBFs. If structures have no significant topological properties, a pseudo-barcode is introduced to insure a better comparison. Moreover, a multiscale biomolecular representation is introduced through the multiscale rigidity function. It naturally induces a multiscale persistent similarity. The multiscale persistent similarity enables an objective-oriented comparison. State differently, it facilitates the comparison of structures in any particular scale of interest. Finally, the proposed method is validated by four different cases. It is found that the persistent similarity can be used to describe the intrinsic similarities and differences between the structures very well.

\end{abstract}

\newpage

{\setcounter{tocdepth}{5} \tableofcontents}

\newpage

\section{Introduction}

The most prominent feature of biological sciences in the $21^{st}$ century is its transition from an empirical, qualitative and phenomenological discipline to a comprehensive, quantitative and predictive one. With the accumulation of gigantic structure and sequence data in Protein Data Bank, Gene Bank, and protein structure classification databanks CATH and SCOP, revolutionary opportunities have arisen for data-driven advances in biological research. An essential component of quantitative biology is geometric analysis. Geometric measurements, algorithms and modeling offer a basis for molecular visualization, bridge the gap between experimental data from X-ray, NMR, and Cryo-electron microscopy, and theoretical models, and play a fundamental role in the analysis of biolmolecular structure, function, dynamics, and transport. Especially with the aid from increasingly powerful high performance computers, geometric analysis becomes more and more deeply involved in biological sciences. However, geometric invariants usually describe local features, such as distances, angles, curvatures, convexity, etc. As a consequence, geometric analysis tends to involve excessive irrelevant structure details and become computationally intractable, especially for macroproteins and protein complexes. Great promises come from a newly founded area in big data analysis, known as topological data analysis (TDA). The essence of TDA is to employ concepts and algorithms from algebraic topology and computational topology to extract or identify intrinsic properties of the data. These intrinsic properties are topological invariants, which describe global features of the structure and are consistent under continuous deformation.

One of the most important tool in TAD is persistent homology, which is a multiscale representation of topological features \cite{Edelsbrunner:2002,Zomorodian:2005,Zomorodian:2008}.  Different from the traditional topological method, persistent homology is able to embed a geometric measurement into topological invariants, thus provides a bridge between geometry and topology. To achieve this, a filtration process is employed in persistent homology. Through the systematical variation of filtration value, a series of continuous topological spaces are generated. Topological invariants like connected components, circles, rings, void and cavities are generated. Their lifespans or persistent times are measured and used as geometric measurements of these topological properties. Historically, the general form of persistent homology is proposed by Robins~\cite{Robins:1999}, Edelsbrunner et al.~\cite{Edelsbrunner:2002}, and Zomorodian and Carlsson \cite{Zomorodian:2005}, independently. Various softwares, including JavaPlex \cite{javaPlex}, Perseus  \cite{Perseus}, Dipha \cite{Dipha}, Dionysus \cite{Dionysus}, jHoles \cite{Binchi:2014jholes}, etc, have been proposed\cite{Bubenik:2007, edelsbrunner:2010,Dey:2008,Dey:2013,Mischaikow:2013}, together with visualization methods, including persistent diagram\cite{Mischaikow:2013}, persistent barcode\cite{Ghrist:2008barcodes}, and persistent landscape\cite{Bubenik:2015statistical}. As a method deeply rooted in algebraic topology, persistent homology has demonstrates great potential in data simplification and complexity reduction \cite{Edelsbrunner:2002,Zomorodian:2005}. It provides new opportunities for researchers from mathematics, computer sciences, computational biology, biomathematics, engineering, etc. Persistent homology has been used in a variety of fields, including shape recognition \cite{DiFabio:2011},network structure \cite{Silva:2005,LeeH:2012,Horak:2009},image analysis \cite{Carlsson:2008,Pachauri:2011,Singh:2008,Bendich:2010,Frosini:2013}, data analysis \cite{Carlsson:2009,Niyogi:2011,BeiWang:2011,Rieck:2012,XuLiu:2012},chaotic dynamics verification \cite{Mischaikow:1999,Kaczynski:2004}, computer vision \cite{Singh:2008} and computational biology \cite{Kasson:2007,YaoY:2009, Gameiro:2013}.

Recently, persistent homology has been used in analyzing fullerene molecules, proteins, DNAs and various other biomolecules \cite{KLXia:2014c, KLXia:2015a,BaoWang:2016a}. To quantitatively analyze the biomolecular structures and functions, I have proposed the concept of topological fingerprint, which is defined as the consistent pattern within the barcode that has particular structure implications\cite{KLXia:2014c}.
Further, I have introduced a multiresolution and multidimensional persistent homology \cite{KLXia:2015c,KLXia:2015b}. With the incorporation of a resolution/scale parameter into a specially designed rigidity function, I can deliver a multiscale structure representation that is able to focus on any scale of interest. More importantly, the corresponding persistent homology analysis provides topological information from various scales, and is proved to be very efficient in handling extremely large data from macroproteins or protein complexes. More recently, based on the persistent homology results, I introduce multiscale persistent functions for biomolecular structure characterization\cite{Xia:2017multiscale}. The essential idea is to combine the multiscale rigidity functions with persistent homology analysis, so as to construct a series of multiscale persistent functions, particularly multiscale persistent entropies, for structure characterization and comparison. My method has been successfully used in protein classification test\cite{Xia:2017multiscale}. All these previous results have demonstrated the great potential of topological analysis, especially persistent homology analysis, in biomolecule structure analysis. In this paper, I further propose a persistent similarity for a quantitative comparison between different structures. My persistent similarity is based on the persistent Betti function (PBF), one type of the persistent functions proposed in my previous work. Different from all the previous structure similarity definitions, my persistent similarity uses only topological persistence information. In this way, it is free from structure or gene-sequence alignment, thus computationally much more efficient, particularly when many structures are considered simultaneously.

Biomolecular structure comparison can not only reveal evolutionary relationships, but also provide insights about biological functional properties. Various definitions of structure and sequence similarity are proposed and are widely used in comparison\cite{koehl:2001protein}. However, all these method involves superposition or alignment at either global scale or common subregions. Computationally, this process is time-consuming and highly inefficient when many structures are involved. Dramatically different from all these methods, my persistent similarity is based on topological characterization, thus free from structure or sequence alignment. More specifically, for each biomolecular structure, I can generate its topological representation, i.e., a series of barcodes. From these barcodes, I can define unique persistent Betti functions. These are very simply one-dimensional continuous functions defined in exactly the same computational domain. In this way, evaluation of similarity between different structures is transferred into the comparison between one-dimensional functions. More importantly, since each biomolecular structure is associated with a unique barcode representation thus a unique set of one-dimensional PBFs, the comparison among various structures becomes much more efficient, as I only need to deal with similarity of one-dimensional PBFs. More importantly, I have introduced the multiscale persistent similarity, so that the structure comparison can be done in the any particular scale of interest.

It has been noticed that a similar ``topological similarity" has been proposed for structure comparison very recently\cite{mate:2014topological,feinauer:2013zinc,mate:2014statistical}. In this model, structure similarity is directly measured from the persistent barcodes, which are generated from persistent homology. Even though the ``topological similarities" \cite{mate:2014topological,feinauer:2013zinc,mate:2014statistical} is also based on persistent homology analysis, my persistence similarity differs greatly from it in several aspects. Firstly, I use the previously proposed PBFs \cite{Xia:2017multiscale}. These PBFs provide a unique representation of persistent barcode. State differently, there is a one to one relation between my PBFs and barcode representations. With these functions, the comparison between different barcodes becomes much more straightforward and efficient. Secondly, a multiscale persistent similarity is defined so that I can systematically compare the structure properties from various scales. Biomolecules, particularly macroproteins or protein complexes, are usually of multiscales ranging from atom, residue, secondary structure, domain, protein monomer, etc. Different topological properties can be found in different scales. And pinpointing to the right scale is of great importance for similarity comparison. In my model, a multiscale rigidity function is employed to represent the structures from various scales. And persistent similarities derived from it capture similarity information in different scales. Thirdly, I introduce a pseudo-barcode to deliver a more precise comparison in the special situation when structures have no significant topological properties. For instance, if a structure has no $\beta_1$ barcodes while the others have, topological similarity between this structure and all the others will always be zero, no matter how long or how many $\beta_1$ barcodes the others have. This ambiguity is avoid by the introduction of a pseudo-barcode in my persistent similarity. 
Fourthly, I introduce weight functions and kernel scales in my PBFs. These parameters give us more flexibility in defining the ``significance" of the bars. It is found that for some biomolecular functions and properties, only some special barcodes matter while the others are irrelevant. And in this situation, my model can play important role. It should be noticed that I deliberately avoid using the term of ``topological similarity", because ``topological similarity" is widely used in network modeling\cite{erten2011:vavien,lei2013:novel}. The term ``persistent similarity" captures the essence of the method and is consistent with all previous notations including, persistent homology, persistent Betti number, persistent entropy, etc. Therefore, I believe it is a much better term to use.

The paper is organized as following. Section \ref{sec:Methodology} is devoted for the introduction of the methodology. I will introduce the persistent homology, define the persistent Betti function and propose the persistent similarity. Further, after the introduction of multiscale persistent homology, I will generalize the persistent similarity to multiscale persistent similarity. Section \ref{sec:results} is dedicated to basic results and discussion. Four different cases are studied, including two similar nucleotide kinases, a series of NMR structures, configurations from molecular dynamics simulation and fullerene $C_{44}$ isomers. The paper ends with a conclusion.

\section{Methodology}\label{sec:Methodology}
The quantitative analysis of similarities between related three-dimensional biomolecular structures is of great importance to structure biology.
The structure relationships not only reveal the biomolecular revolutionary connections, but also help with understanding of biomolecular functional properties and further the development of new and improved materials. Persistent similarity is proposed in the current paper to facilitate an efficient quantitative comparison of biomolecular structures. Since the persistent homology analysis provides a delicate balance between topological simplification and geometric details, the persistent similarity derived from it captures the intrinsic structure properties. To be more specific,
the results from the persistent homology analysis are uniquely represented by the proposed persistent Betti functions (PBFs). In this way, the description of biomolecular structures is dramatically simplified into several one-dimensional PBFs. And the comparison of these PBFs results in the persistent similarity. Further, using the multiscale rigidity functions, a multiscale biomolecular representation is delivered and naturally induces a multiscale persistent similarity. A detailed description is given below.

\subsection{Persistent homology}\label{sec:PHA}

\begin{figure}
\begin{center}
\begin{tabular}{c}
\includegraphics[width=0.95\textwidth]{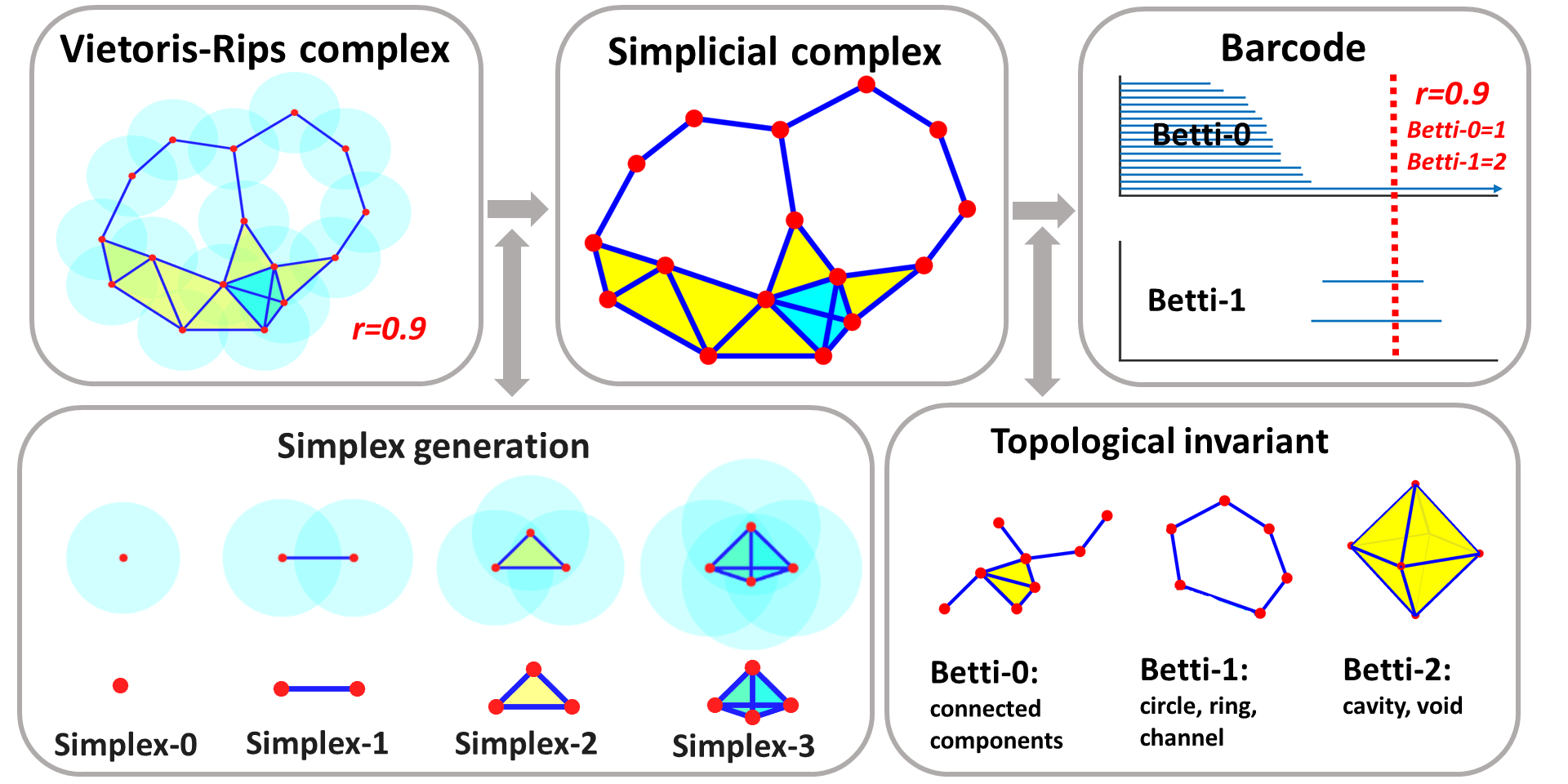}
\end{tabular}
\end{center}
\caption{ The illustration of the basic concepts in persistent homology analysis. A more general filtration process is demonstrated in Figure \ref{fig:filtration}. The Vietoris-Rips complex is generated with sphere radius equal to 0.9. Its topological information is indicated by the red dash line in the barcode.
}
\label{fig:top}
\end{figure}

Since persistent similarity is derived from persistent homology, a detailed introduction of persistent homology is given in this section. To avoid the heavy mathematical notations and present essential ideas more straightforwardly, I will only focus on the simplical complex with direct geometric implications. Further, the homology is calculated in $Z_2$ field and only Vietoris-Rips complex is considered for simplicial complices generation. Interested readers are referred to more detailed description in papers\cite{Edelsbrunner:2002,Zomorodian:2005,Zomorodian:2008}.

\begin{figure}
\begin{center}
\begin{tabular}{c}
\includegraphics[width=0.8\textwidth]{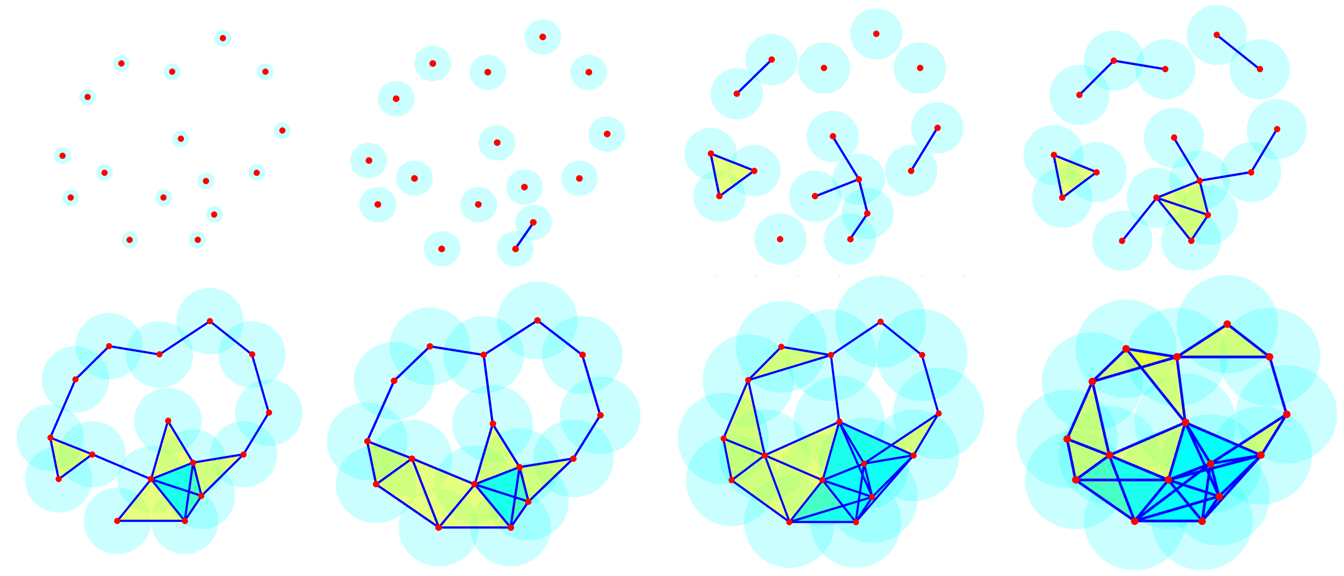}
\end{tabular}
\end{center}
\caption{ The filtration process and barcode representation of a fullerence $C_{60}$ molecule. During the filtration process, each Carbon atom is associated with a sphere, whose radius increases systematically  to generate topologies in various scales. In the molecular barcode representation, each bar represents a homology generator and has unique chemical or physical properties. For $\beta_0$ bars, they are related to atom bonds. The ring structures are represented by $\beta_1$ bars. The void or cavity structures are captured in $\beta_2$ bars.
}
\label{fig:filtration}
\end{figure}

Simply speaking, homology is a mathematical representation of topological invariants, such as connected components, circle, rings, channels, cavity, void, cage, etc. Persistent homology gives a geometric measurement, i.e., a size, to these invariants. Figure \ref{fig:top} and Figure \ref{fig:filtration} illustrate the essential concepts used in persistent homology.

\paragraph{Simplicial complex}
Simplices are the build block for simplicial complex. A set of $k+1$ affine independent points $v_0,v_1,v_2,\cdots,v_k$ can form a $k$-simplex $\sigma^k=\{v_0,v_1,v_2,\cdots,v_k\}$ as following,
\begin{eqnarray}\label{eq:couple_matrix1}
\sigma^k=\left\{\lambda_0 v_0+\lambda_1 v_1+ \cdots +\lambda_k v_k \mid \sum^{k}_{i=0}\lambda_i=1;0\leq \lambda_i \leq 1,i=0,1, \cdots,k \right\}.
\end{eqnarray}
Geometrically, a 0-simplex is a vertex, a 1-simplex is an edge, a 2-simplex is a triangle, and a 3-simplex represents a tetrahedron, just as depicted in Figure \ref{fig:top}. The $i$-dimensional face of $\sigma^k$ is the convex hull formed by $i+1$ vertices from $\sigma^k$ ($k>i$). A simplicial complex $K$ is a finite set of simplices that satisfy two essential conditions, i.e., 1) any face of a simplex from  $K$  is also in  $K$; 2) the intersection of any two simplices in  $K$ is either empty or shares faces. Further, we denote an oriented $k$-simplex as $[\sigma^k]$. An oriented simplex is a simplex together with an orientation, i.e., ordering of its vertex set.

\paragraph{Homology}
A $k$-chain $c$ is a linear combination of $k$-simplexes $c=\sum_{i}\alpha_i\sigma^k_i$ with $\{ \alpha_i \in Z_2 \} $. An Abelian group $C_k(K, \mathbb{Z}_2)$ is formed by the set of all $k$-chains from the simplicial complex $K$ together with addition operation (modulo-2). A boundary operator $\partial_k$ is defined as $\partial_k: C_k \rightarrow C_{k-1}$. The boundary of {an oriented $k$-simplex} $[\sigma^k]=[v_0,v_1,v_2,\cdots,v_k]$ can be denoted as,
\begin{eqnarray}
\partial_k [\sigma^k] = \sum^{k}_{i=0} [ v_0, v_1, v_2, \cdots, \hat{v_i}, \cdots, v_k ].
\end{eqnarray}
Here $[v_0, v_1, v_2, \cdots ,\hat{v_i}, \cdots, v_k ]$ means a $(k-1)$ oriented simplex, which is generated by the elimination of vertex $v_i$. Further, one has $\partial_0= 0$ and $\partial_{k-1}\partial_k= 0$. The $k$-th cycle group $Z_k$ and the $k$-th boundary group $B_k$ are the subgroups of $C_k$ and can be defined as,
\begin{eqnarray}
&& Z_k={\rm Ker}~ \partial_k=\{c\in C_k \mid \partial_k c=0\}, \\
&&  B_k={\rm Im} ~\partial_{k+1}= \{ c\in C_k \mid \exists d \in C_{k+1}: c=\partial_{k+1} d\}.
\end{eqnarray}
Their elements are called the $k$-th cycle and the $k$-th boundary, respectively. It can be noticed that $B_k\subseteq Z_k$, as the boundary of a boundary is always zero $\partial_{k-1}\partial_k= 0$.
The $k$-th homology group $H_k$ is the quotient group generated by the $k$-th cycle group $Z_k$ and $k$-th boundary group $B_k$: $H_k=Z_k/B_k$. The rank of $k$-th homology group is called $k$-th Betti number and it can be calculated by
\begin{eqnarray}
\beta_k = {\rm rank} ~H_k= {\rm rank }~ Z_k - {\rm rank}~ B_k.
\end{eqnarray}
As indicated in Figure \ref{fig:top}, the geometric meanings of Betti numbers in $\mathbb{R}^3$ are as following: $\beta_0$ represents the number of isolated components; $\beta_1$ is the number of one-dimensional loops, circles, or tunnels; $\beta_2$ describes the number of two-dimensional voids or holes. Together, the Betti number sequence { $\{\beta_0,\beta_1,\beta_2\}$} describes the intrinsic topological property of the system.

\paragraph{Rips complex}
For a point set $X \in \mathbb{ R}^N$, one defines a cover of closed balls centered at $x$ with radius $\epsilon$. A Rips simplex (or Vietoris-Rips simplex) $\sigma$ is generated if the largest distance between any of its vertices reaches $2\epsilon$. Figure \ref{fig:top} illustrates the generation of the Rips simplex.

\paragraph{Filtration}
In the generation of Rips complex, a radius parameter $\epsilon$ is used. However, how to find the best suitable $\epsilon$ so that it can best capture the underling space has been long standing problem. To solve this problem, the idea of filtration has been proposed. As illustrated in Figure \ref{fig:filtration}, instead of finding the best suitable value, an ever-increasing $\epsilon$ value is used to generate a series of topological spaces. In this way, the associated topological invariants will have certain ``lifespan", that is some topological invariants last for a wide range of $\epsilon$ values, but some invariants disappear very quickly when $\epsilon$ value changes. 

\paragraph{Persistent homology}
The filtration can be described as a nested sequence of its subcomplexes,
\begin{eqnarray}
\varnothing = K^0 \subseteq K^1 \subseteq \cdots \subseteq K^m=K.
\end{eqnarray}
And the $p$-persistent $k$-th homology group at filtration time $i$ can be represented as
\begin{eqnarray}
H^{i,p}_k=Z^i_k/(B_k^{i+p}\bigcap Z^i_k).
\end{eqnarray}
Essentially, persistence gives a geometric measurement of the topological invariant.

\subsection{Persistent similarity}\label{sec:MPF}
\begin{figure}
\begin{center}
\begin{tabular}{c}
\includegraphics[width=0.98\textwidth]{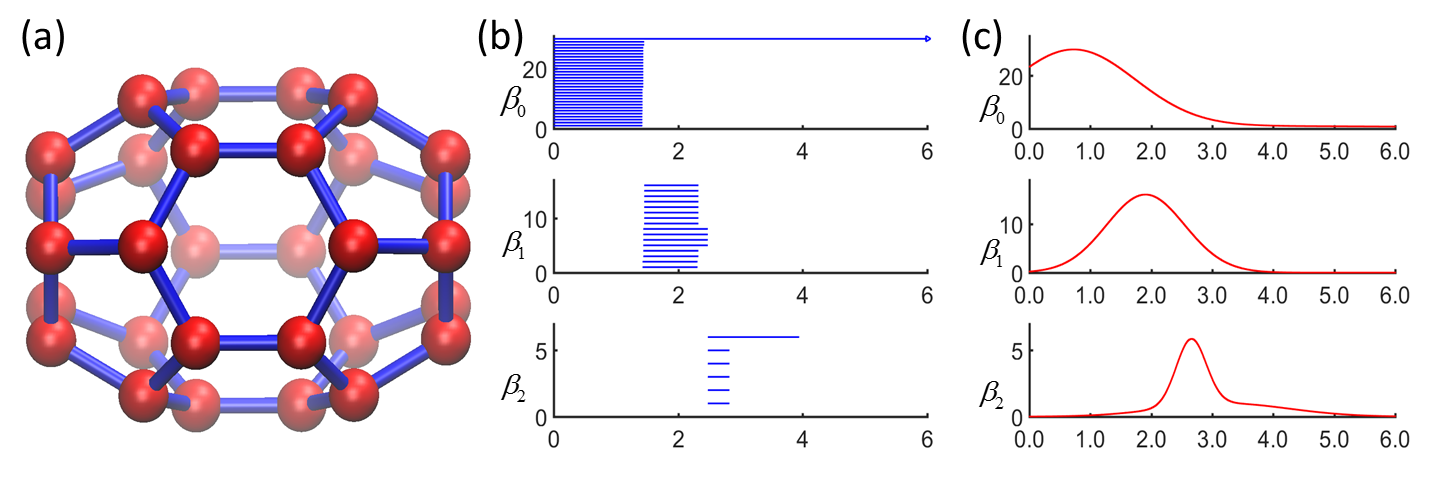}
\end{tabular}
\end{center}
\caption{ Fullerence $C_{44}$ molecule structure, barcode representation and PBFs. ({\bf a}) The cage structure of fullerence $C_{44}$. ({\bf b}) The barcode representation. Each bar represents a homology generator and has unique chemical or physical properties. For $\beta_0$ bars, they are related to atom bonds. The pentagon and hexagon ring structures are represented by $\beta_1$ bars. Its cavity or cage structure are captured in $\beta_2$ bars. ({\bf c}) Fullerence $C_{44}$ PBFs.
}
\label{fig:c30}
\end{figure}

The results from the persistent homology analysis can be represented as following,
\begin{eqnarray}
\{ L_{k,j}=[a_{k,j}, b_{k,j}] | k=0,1,2; j=1,2,3,....,N_k \},
\end{eqnarray}
where parameter $k$ is the dimension of Betti number $\beta_k$, parameter $j$ indicates the $j$-th homology generator and $N_k$ is the number of $\beta_k$ generator.

Further, I define the sets of barcodes in the $k$-th dimension,
$$ L_{k}= \{ L_{k,j}, j=1,2,3,....,N_k\}, \quad k=0, 1, 2.$$

To visualize the persistent homology results, I use the barcode plot as illustrated in Figure \ref{fig:c30} ({\bf b}). For fullerence $C_{44}$ barcodes, the length of $\beta_0$ bars is the atom bond length. The number of $\beta_0$ bars are the total number of atoms in the molecular. Further, the pentagon and hexagon ring structures are represented by $\beta_1$ bars. The cavity or cage structure of fullerene $C_{44}$ is captured in $\beta_2$ bars. These basic chemical implications of the barcodes are very consistent. Moreover, it should be noticed that there is no general way of defining the sequence of the barcodes. I simply define the sequence by birth times of bars.

\paragraph{Persistent Betti function (PBF)}
Based on the barcode, I can build up various model to further explore the biomolecular structure, flexibility, function and dynamics\cite{KLXia:2015c,Xia:2017multiscale}. Among them, persistent Betti function is defined as,
\begin{eqnarray}
f(x;L_{k})= \sum_{j} w_{k,j} e^{-\left(\frac{x- \frac{b_{k,j}+a_{k,j}}{2}}{\sigma (b_{k,j}-a_{k,j})}\right)^\kappa},  \quad \kappa >0; k=0, 1, 2.
\end{eqnarray}
where $w_{k,j}$ is the weight function for the $j$-th barcode of $\beta_k$. Parameter $\sigma$ is the resolution parameter. It should be noticed that even thought there is no meaningful sequence arrangement for barcodes, barcodes derived from the molecules are highly organized. Each bar or each type of bars has its unique structural, physical and chemical implication. With this consideration, I can assign or define a weight value $w_{k,j}$ to each bar or each type of bars. Normally, the weight function and resolution parameter are all chosen as 1, i.e., $\sigma=1$ and $w_{k,j}=1$ for all $k$ and $j$.

More interestingly, PBF provides a unique transformation of persistent barcodes into 1D continuous functions. There is a strict one-to-one correlation between barcodes and PBFs. In this way, any complicated biomolecular structures can be uniquely represented by three 1D PBFs, which dramatically reduced the dimensionality and complexity.

\paragraph{Persistent similarity}
\begin{figure}
\begin{center}
\begin{tabular}{c}
\includegraphics[width=0.8\textwidth]{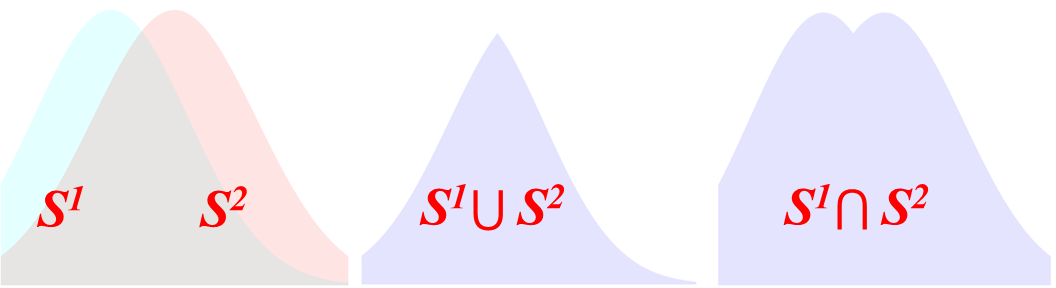}
\end{tabular}
\end{center}
\caption{ To define persistent similarity from persistent functions. Capitals $S^1$ and $S^2$ represent regions under two persistent functions. The persistent similarity $P$ is defined to be the quotient between the intersect area and the union area, i.e., $P=\frac{\int S^1 \bigcap S^2}{\int S^1 \bigcup S^2}$
}
\label{fig:similarity}
\end{figure}

For two different biomolecular structures denoted as $F_1$ and $F_2$, if their PBFs are $f_1(x;L_{k})$ and $f_2(x;L_{k})$ and the regions below the two functions is defined as $S^1_k=\{(x,y) | 0 \leq y(x) \leq f_1(x;L_k); 0 \leq x \leq r_f\}$ and $S^2_k=\{(x,y) | 0 \leq y(x) \leq f_2(x;L_k); 0 \leq x \leq r_f\}$,  the persistent similarity can be defined as
\begin{eqnarray}
P_k(F_1,F_2)=\frac{\int S^1_k \bigcap S^2_k }{\int S^1_k \bigcup S^2_k}, \quad k=0, 1, 2  .
\end{eqnarray}
Here $r_f$ is the filtration size. A suitable filtration size is not unique. Usually, it is chosen as the smallest value, after which the filtration generates no significant topological properties.

The above definition of persistent similarity is equivalent to,
\begin{eqnarray}
P_k(F_1,F_2)=\frac{\int \min \{f_1(x;L_{k}),f_2(x;L_{k})\}}{\int \max \{f_1(x;L_{k}),f_2(x;L_{k})\}}, \quad k=0, 1, 2.
\end{eqnarray}

Some structures may not have certain significant topological properties. For instance, small molecules may not have $\beta_1$ barcodes. In this situation, if it is compared with other molecules with $\beta_1$ barcodes, the similarity is always zero, no matter how many or how long the $\beta_1$ barcodes in the other structures. This ambiguity brings trouble in structure comparison. To overcome this problem, I introduce a pseudo-barcode into the PBFs and definite a new PBF as,
\begin{eqnarray}
f^{pseudo}(x;L_{k})= w_{k,0}+\sum_{j} w_{k,j} e^{-\left(\frac{x- \frac{b_{k,j}+a_{k,j}}{2}}{\sigma (b_{k,j}-a_{k,j})}\right)^\kappa},  \quad \kappa >0; k=0, 1, 2.
\end{eqnarray}
Essentially, a small weight value $w_{k,0}$ is introduced to avoid the situation when PBF is zero function. In this way, I can reduce the ambiguity.

\subsection{Multiscale persistent similarity}\label{sec:MPS}
Biomolecular data are usually highly complicated and essentially multiscale. To overcome this challenge, I have proposed a multiresolution/multiscale persistent homology\cite{Xia2015:multiresolution}. The essential idea is to match the scale of interest with appropriate resolution in the topological analysis. Simply speaking, a resolution parameter is introduced into my multiscale rigidity function, and by turning this parameter, I can focus my topological analysis on any interesting scales. The multiscale rigidity function, which is derived from flexibility and rigidity index (FRI) method\cite{KLXia:2013d,Opron:2014,Opron:2015communication,Opron:2014b, Xia:2015multiscale,Nguyen:2016generalized}, is key to the multiscale persistent homology. By using this function, I can convert a discrete point cloud data into a series of continuous density functions. The conversion is realized by using a kernel function with a resolution or scale parameter. And this special parameter enables us to facilitate a multiscale analysis of complex data. More details will be discussed below.

\paragraph{Multiscale rigidity function}\label{sec:Multiscale_rigidity_function}
For a data set with a total $N$ entries, which can be physical elements like atoms, residues and domains or data components like points, pixels and voxels, if one assumes their generalized coordinates are ${\bf r}_1, {\bf r}_2,\cdots, {\bf r}_N$, a multiscale rigidity function of the data can be expressed as,
\begin{eqnarray}\label{eq:rigidity_function}
\mu({\bf r},\eta)=\sum_{j}^N w_j\Phi(\parallel {\bf r}- {\bf r}_{j}\parallel;\eta)
\end{eqnarray}
where $w_j$ is a weight, which usually is chosen as the atomic number, for example, its value is 6 for Carbon atom and 8 for Oxygen atom. The parameter $\eta$ is the resolution or scale parameter. The function $\Phi(\parallel {\bf r}- {\bf r}_{j}\parallel;\eta)$ is a kernel function. Commonly used kernel functions are  generalized exponential functions,
\begin{eqnarray}\label{eq:couple_matrix1}
\Phi(\parallel {\bf r}- {\bf r}_{j}\parallel ;\eta, \kappa) =    e^{-\left( \parallel {\bf r}- {\bf r}_{j}\parallel /\eta \right)^\kappa},    \quad \kappa >0
\end{eqnarray}
or generalized Lorentz functions,
\begin{eqnarray}\label{eq:couple_matrix2}
 \Phi(\parallel {\bf r}- {\bf r}_{j}\parallel;\eta, \upsilon) =  \frac{1}{1+ \left( \parallel {\bf r}- {\bf r}_{j}\parallel /\eta \right)^{\upsilon}},  \quad  \upsilon >0.
 \end{eqnarray}
It can be noticed that the larger the $\eta$ value, the lower the resolution is. A multiscale geometric model can be naturally derived from my multiscale rigidity functions. An example is given in Section \ref{sec:results} Case 4.

\paragraph{Multiscale persistent homology}\label{sec:Multiscale_PH}
Based on the multiscale rigidity function, I have proposed multiscale persistent homology\cite{Xia2015:multiresolution,KLXia:2015d}. In this model, I linearly rescale all the rigidity function values to the region $[0, 1]$ using formula
\begin{eqnarray}\label{eq:scaled_rigidity_function}
\mu^s({\bf r},\eta)=1.0-\frac{\mu({\bf r},\eta)}{\mu_{\max}(\eta)}.
\end{eqnarray}
Here $\mu({\bf r},\eta)$ and $\mu^s({\bf r},\eta)$ are the original and normalized rigidity function, and $\mu_{\max}(\eta)$ is the maximum value of the original rigidity function.

I can perform the persistent homology analysis on the rescale rigidity functions. For density data, the filtration parameter is chosen as the isovalue or isosurface value. To be more specific, for each density isovalue, I can generate a molecular surface. With the continuous variation of this value, a series of molecular surfaces are generated. Molecular surfaces are topological spaces and the their homology information can be calculated from Morse theory \cite{Zomorodian:2005}. Based on these surfaces,  Morse complexes are generated and form a nested sequences. In this way, persistent homology analysis can be employed\cite{Mischaikow:2013,Harker:2010,mischaikow:nanda}.

\paragraph{Multiscale persistent similarity}\label{sec:MPF}
A series of barcodes from various scales are generated in the multiscale persistent homology and can be represented as following,
\begin{eqnarray}
\{L_{k,j}(\eta)=[a_{k,j}(\eta), b_{k,j}(\eta)] | k=0,1,2; j=1,2,3,....,N_k(\eta) \}.
\end{eqnarray}
Similar to the previous definition, parameter $k$ is the dimension of Betti number $\beta_k$, parameter $j$ indicates the $j$-th barcode and $N_k$ is the number of $\beta_k$ barcodes. And the sets of barcodes in the $k$-th dimension is represented as,
$$ L_{k}(\eta)= \{ L_{k,j}(\eta), j=1,2,3,....,N_k(\eta)\}, \quad k=0,1,2$$

Further the multiscale persistent Betti function is represented as,
\begin{eqnarray}
f(x;L_{k}(\eta))= \sum_{j} w_{k,j}(\eta) e^{-\left(\frac{x- \frac{b_{k,j}(\eta)+a_{k,j}(\eta)}{2}}{\sigma(\eta) (b_{k,j}(\eta)-a_{k,j}(\eta))}\right)^\kappa},  \quad \kappa >0, k=0,1,2.
\end{eqnarray}
Again $w_{k,j}(\eta)$ is the weight function for the $j$-th barcode of $\beta_k$. Parameter $\sigma(\eta)$ is the resolution or scale parameter.

In this same way, the multiscale persistent similarity between structures $F_1(\eta)$ and $F_2(\eta)$ can be defined as
\begin{eqnarray}
P_k(F_1(\eta),F_2(\eta))=\frac{\int \min \{f_1(x;L_{k}(\eta)),f_2(x;L_{k}(\eta))\}}{\int \max \{f_1(x;L_{k}(\eta)),f_2(x;L_{k}(\eta))\}}, \quad k=0, 1, 2.
\end{eqnarray}
The multiscale persistent similarity enable us to compare the structure properties from various scales.

\section{Results and discussion}\label{sec:results}
In this section, I validate my persistent similarity method using four different cases. In the first case, I consider two similar nucleotide kinases 1AKY and 1GKY. I calculate their persistent similarities for both all-atom-without-hydrogen model and $C_{\alpha}$ coarse-grained model. The calculated persistent similarities are in the middle range, indicating some potential similarities between two structures. In the second cases, a series of structures of protein 2KIX from NMR experiment are considered. These configurations are highly consistent with only small thermal fluctuations. The persistent similarities for both $\beta_0$ and $\beta_1$ are of large values, indicating a strong similarity between all the frames. The third cases is devoted to the validation of multiscale persistent similarity. The steered molecular dynamic simulation results of protein Titin are analyzed from two different scales. I find that even thought structurally, four extracted frames differ greatly, they local scale properties bear great similarities, meaning local structures have no significant changes in the simulation. Further, I calculate the persistent similarities based on global scale models, a dramatic reduction of similarity values are observe, which is highly consistent with the general unfolding process. The last case is employed for the study of fullerene $C_{44}$ isomers. Since the total curvature energy is highly related to the longest $\beta_2$ barcode, the persistent similarity is then defined only on this particulary barcode by using the special weight parameters. I establish a linear relation between persistent similarities and total curvature energy differences. More importantly, I find that this linear relation is particularly strong when I use extreme structures for comparison. To void confusion, in all four cases except the last one, the weight function and resolution parameter in the PBF are all chosen as 1, i.e., $\sigma=1$ and $w_{k,j}=1$ for all $k$ and $j$.


\subsection{Case 1: Two similar nucleotide kinases}

\begin{figure}
\begin{center}
\begin{tabular}{c}
\includegraphics[width=0.5\textwidth]{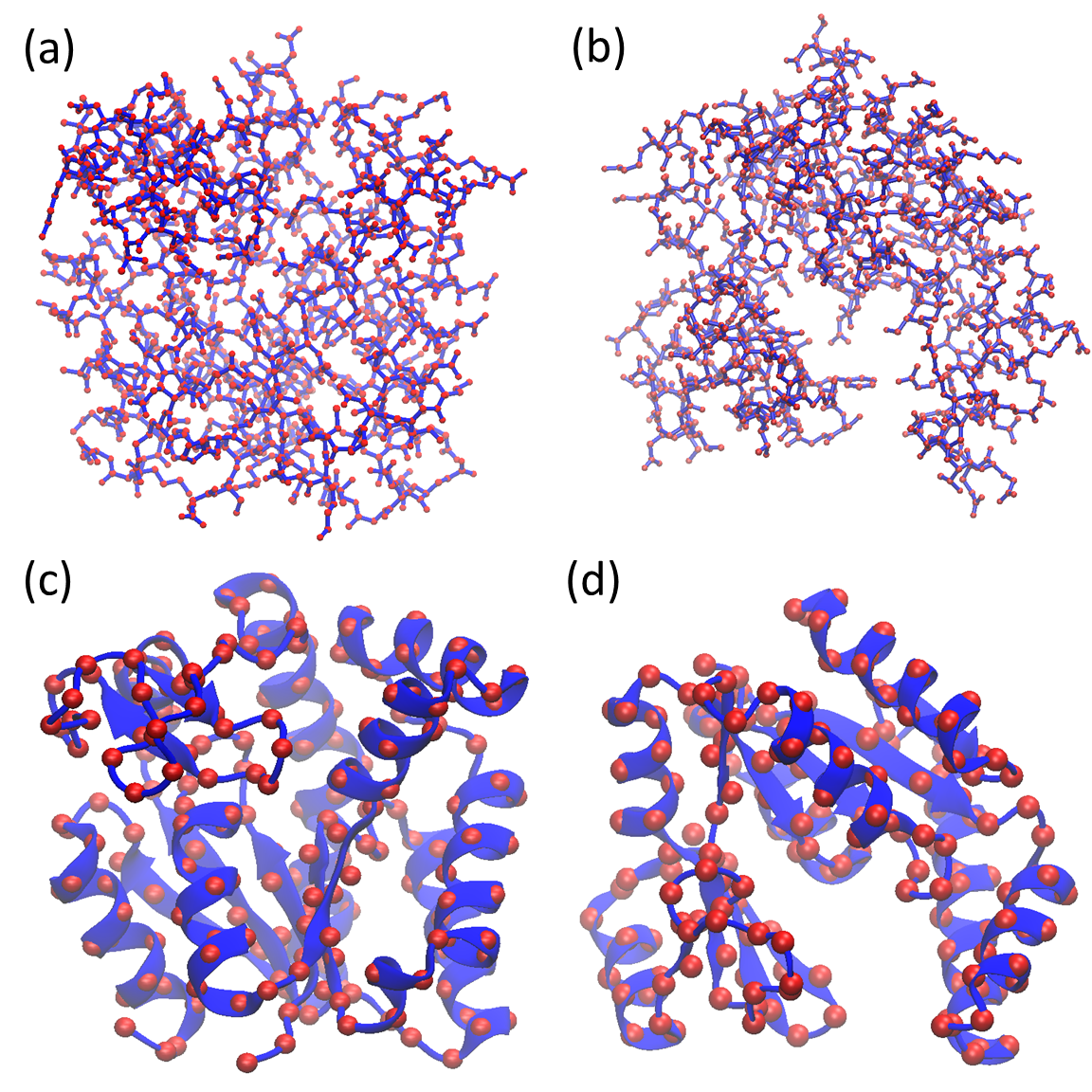}
\end{tabular}
\end{center}
\caption{ The biomolecular structures of Kinase proteins 1AKY and 1GKY. Two different representations are employed, i.e., all-atom-without-hydrogen model and $C_{\alpha}$ coarse-grained model. ({\bf a}) and ({\bf b}) are all-atom except hydrogen models of 1AKY and 1GKY, respectively.  ({\bf c}) and ({\bf d}) are $C_{\alpha}$ coarse-grained model of 1AKY and 1GKY, respectively. }
\label{fig:1aky_homology}
\end{figure}

\begin{figure}
\begin{center}
\begin{tabular}{c}
\includegraphics[width=0.6\textwidth]{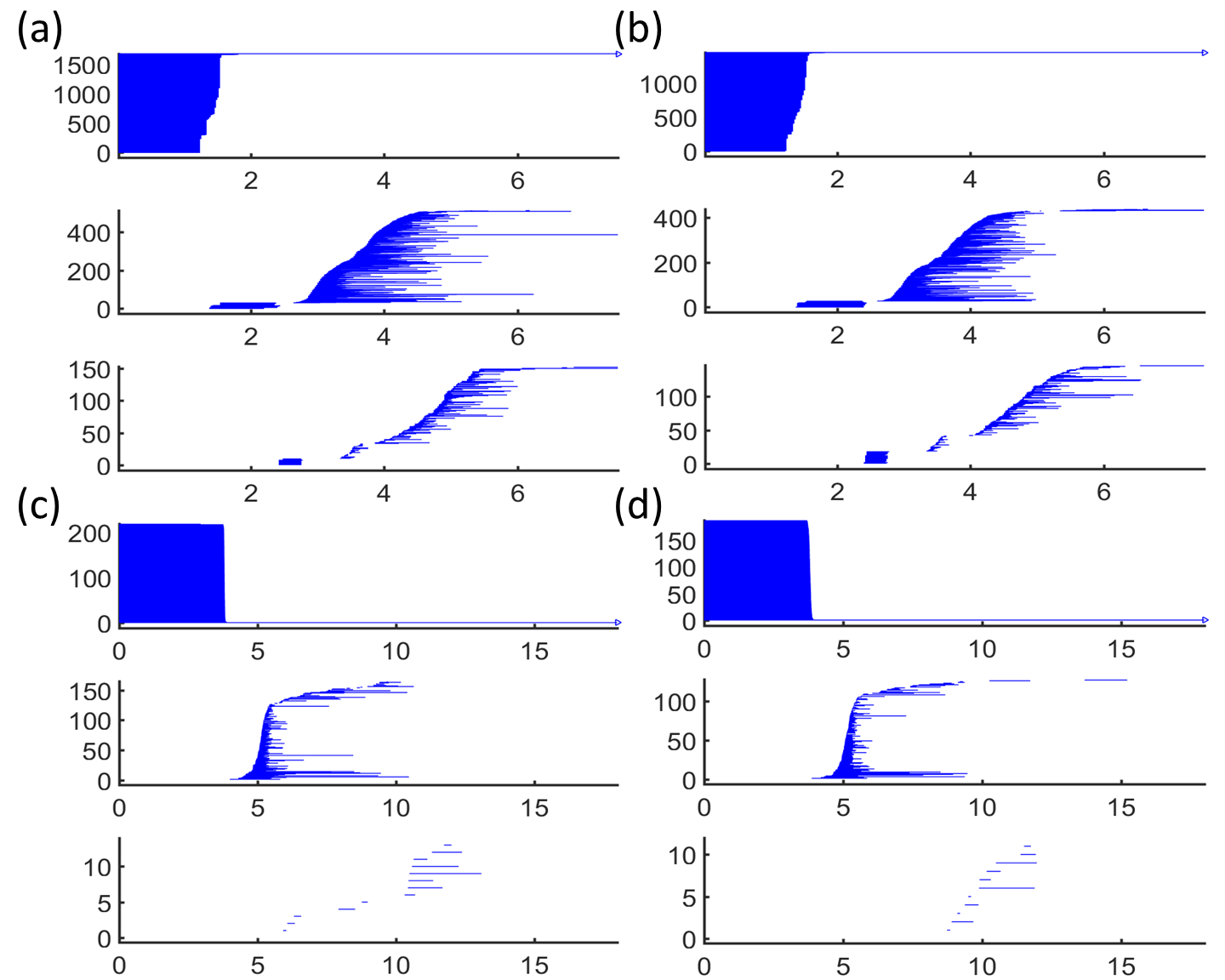}
\end{tabular}
\end{center}
\caption{Persistent barcodes for Kinase proteins 1AKY and 1GKY in different representations. ({\bf a}) and ({\bf b}) are barcode representations for all-atom-without-hydrogen models of 1AKY and 1GKY, respectively.  ({\bf c}) and ({\bf d}) are barcode representations of $C_{\alpha}$ coarse-grained model of 1AKY and 1GKY, respectively. }
\label{fig:1aky_barcodes}
\end{figure}

\begin{figure}
\begin{center}
\begin{tabular}{c}
\includegraphics[width=0.6\textwidth]{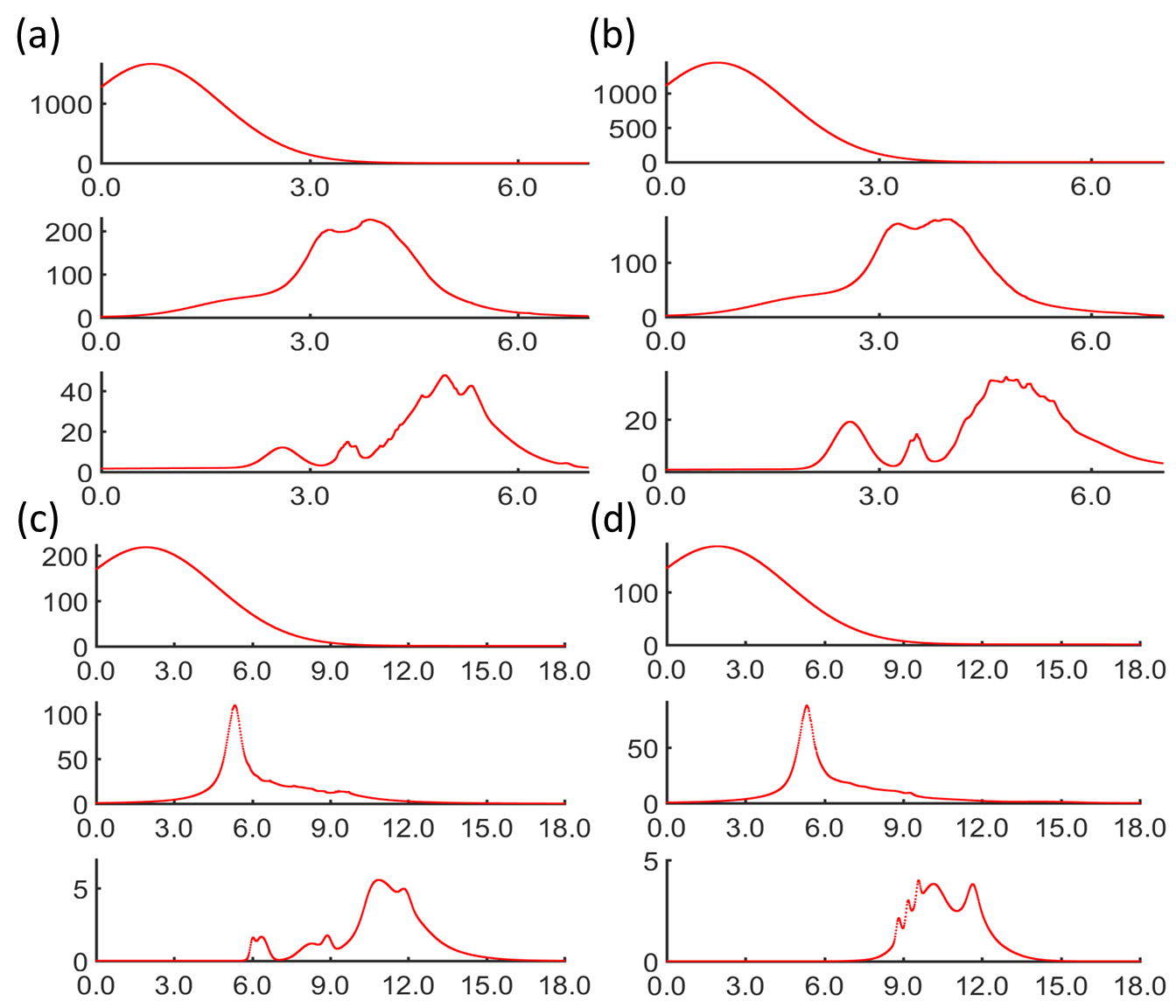}
\end{tabular}
\end{center}
\caption{ PBFs for Kinase proteins 1AKY and 1GKY in different representations. ({\bf a}) and ({\bf b}) are PBFs derived from all-atom-without-hydrogen models of 1AKY and 1GKY, respectively.  ({\bf c}) and ({\bf d}) are PBFs derived from $C_{\alpha}$ coarse-grained model of 1AKY and 1GKY, respectively. The persistence similarities between ({\bf a}) and ({\bf b}) are 0.860, 0.819, 0.795 for $\beta_0$, $\beta_1$ and $\beta_2$, respectively. The persistence similarities between ({\bf c}) and ({\bf d}) are 0.857, 0.770, 0.485 for $\beta_0$, $\beta_1$ and $\beta_2$, respectively.}
\label{fig:1aky_pbf}
\end{figure}

In the first case, I consider two nucleotide kinases (1AKY and 1GKY) used in structural alignment \cite{gelly:2011ipba}. Two structure description, i.e., all-atom-without-hydrogen model and $C_{\alpha}$ coarse-grained model, are used for persistent similarity evaluation. Figure \ref{fig:1aky_homology} illustrates the structure properties of these two proteins. It can be seen that these two structures do share some similar structure components, like the $\alpha$-helix on the left boundary regions and the $\beta$-sheets in the middle regions.

To have a more quantitatively understanding of the structural similarity, I generate the barcodes for both structures in both representations. Figure \ref{fig:1aky_barcodes} $(\bf a)$ and $(\bf b)$ are barcodes for all-atom-without-hydrogen model of 1AKY and 1GKY, respectively.  As stated above, the length of short $\beta_0$ bars represents atomic bond length. And the number of $\beta_0$ bars is the number of atoms in the system. In this way, from $\beta_0$ bars, chemical components of the structure can be understood. More chemical implications can be learned from $\beta_1$ bars. Previously, I have found that, short $\beta_1$ bars located in local region around 2.0 \AA~ represent pentagon and hexagon rings in aromatic residues \cite{KLXia:2014persistent,KLXia:2015c}. Particularly, the hexagon rings can further manifest themselves in local $\beta_2$ bars. The global structure properties captured by $\beta_1$ bars appear much later in the filtration. For all-atom-without-hydrogen model, there is clear separation of local and global type of $\beta_1$ bars in both $\beta_1$ and $\beta_2$ barcodes.

Figure \ref{fig:1aky_barcodes} $(\bf c)$ and $(\bf d)$ are barcodes for $C_{\alpha}$ coarse-grained modeling of 1AKY and 1GKY, respectively. One can see that the length of all short $\beta_0$ bars are around 3.8 \AA,~ i.e., the distance between the two adjacent $C_{\alpha}$ atoms. More over, all three types of barcodes are dramatically reduced. Particularly the $\beta_2$ barcodes.

With these barcode results, I can generate the persistent Betti functions. Figure \ref{fig:1aky_pbf} illustrates the basic pattern of PBFs for two proteins in the same sequence as Figure \ref{fig:1aky_homology} and \ref{fig:1aky_barcodes}. It can be seen that PBFs are simply one-dimensional functions, which can be easily compared with each other. For all-atom-without-hydrogen models, the persistent similarities are 0.860, 0.819, 0.795 for $\beta_0$, $\beta_1$ and $\beta_2$, respectively. For $C_{\alpha}$ coarse-grained model, the persistence similarities are 0.857, 0.770, 0.485 for $\beta_0$, $\beta_1$ and $\beta_2$, respectively. However, in coarse-grained model, there are only a few $\beta_2$ barcodes. And many of these bars are extremely short, meaning they are transient state with no topological significance. Therefore, in $C_{\alpha}$ coarse-grained models, I only consider the $\beta_0$ and $\beta_1$ persistent similarity.

\subsection{Case 2: NMR configurations}

\begin{figure}
\begin{center}
\begin{tabular}{c}
\includegraphics[width=0.9\textwidth]{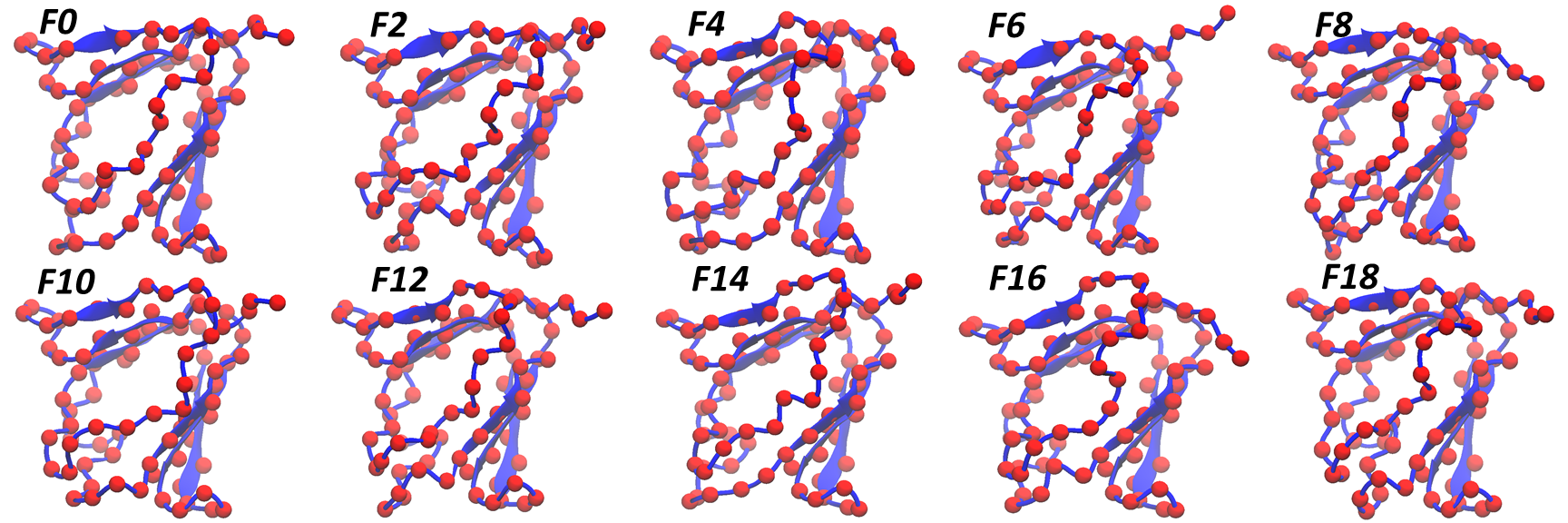}
\end{tabular}
\end{center}
\caption{ The $C_{\alpha}$ coarse-grained representations of protein 2KIX. There are totally twenty configurations (denoted as $F1$ to $F20$)in this pdb data generated from NMR. I take ten different configurations among them. It can be seen that they all have very similar structures.}
\label{fig:2K1X_structure}
\end{figure}

\begin{figure}
\begin{center}
\begin{tabular}{c}
\includegraphics[width=0.8\textwidth]{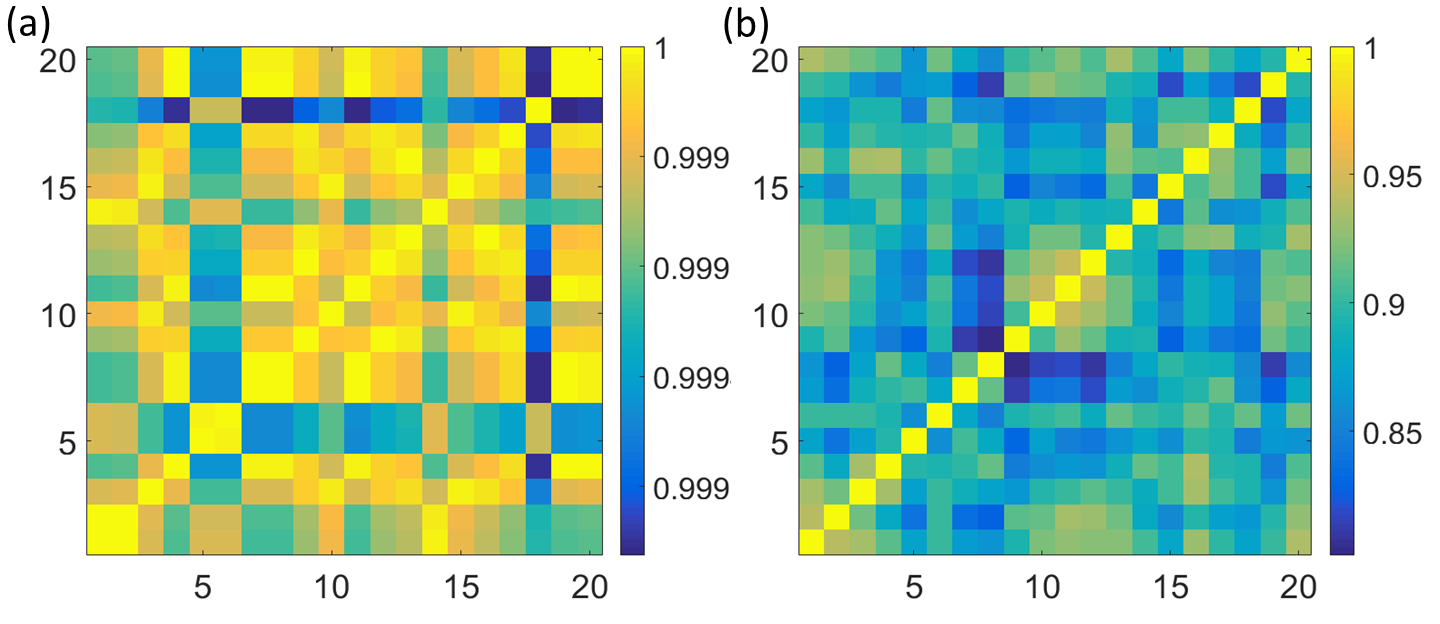}
\end{tabular}
\end{center}
\caption{ The persistent similarities between twenty different NMR configurations of protein 2KIX. ({\bf a}) Betti-0 persistent similarities. ({\bf b}) Betti-1 persistent similarities. It can be seen that Betti-0 persistent similarities are very large and all closed to 1.0.  Betti-1 persistent similarities are also relatively high.}
\label{fig:2K1X_ps}
\end{figure}

In the second case, I consider a NMR solution structure of M-crystallin in calcium free form (PDB ID: 2KIX). There are totally 20 frames in this data. All these configurations are very similar to each other with only small variations due to thermal fluctuation. To quantitatively measure the structure similarity, I consider the $C_{\alpha}$ coarse-grained representation as illustrated in Figure \ref{fig:2K1X_structure}. It should be noticed that I only demonstrate 10 configurations out of 20.

I further perform the persistent homology analysis on these configurations. The persistent similarity is illustrated in Figure \ref{fig:2K1X_ps}. More specifically, Figure \ref{fig:2K1X_ps} ({\bf a}) and ({\bf b}) are persistent similarity for $\beta_0$ and $\beta_1$, respectively. It can be seen that all persistent similarity values are very large. Particularly for $\beta_0$, the values are all around 1.000. For $\beta_1$, the smallest persistent similarity among these twenty configurations is 0.802.

By the comparison of the persistent similarity values between Case 1 and Case 2, one can see that the persistent similarity gives a very reasonable evaluation of the structure similarity. For structures with same chemical components, the $\beta_0$ persistent homology is all around 0.999, indicating the equivalent structure elements. This is also consistent with chemical implications of $\beta_0$ bars. Further, the  $\beta_1$ persistent similarities for 2K1X structures are all larger the similarity between 1AKY and 1GKY. This is also reasonable from the observations, as 2K1X structures show a clear consistence between each other.

\subsection{Case 3: Steered dynamic simulation}

\begin{figure}
\begin{center}
\begin{tabular}{c}
\includegraphics[width=0.8\textwidth]{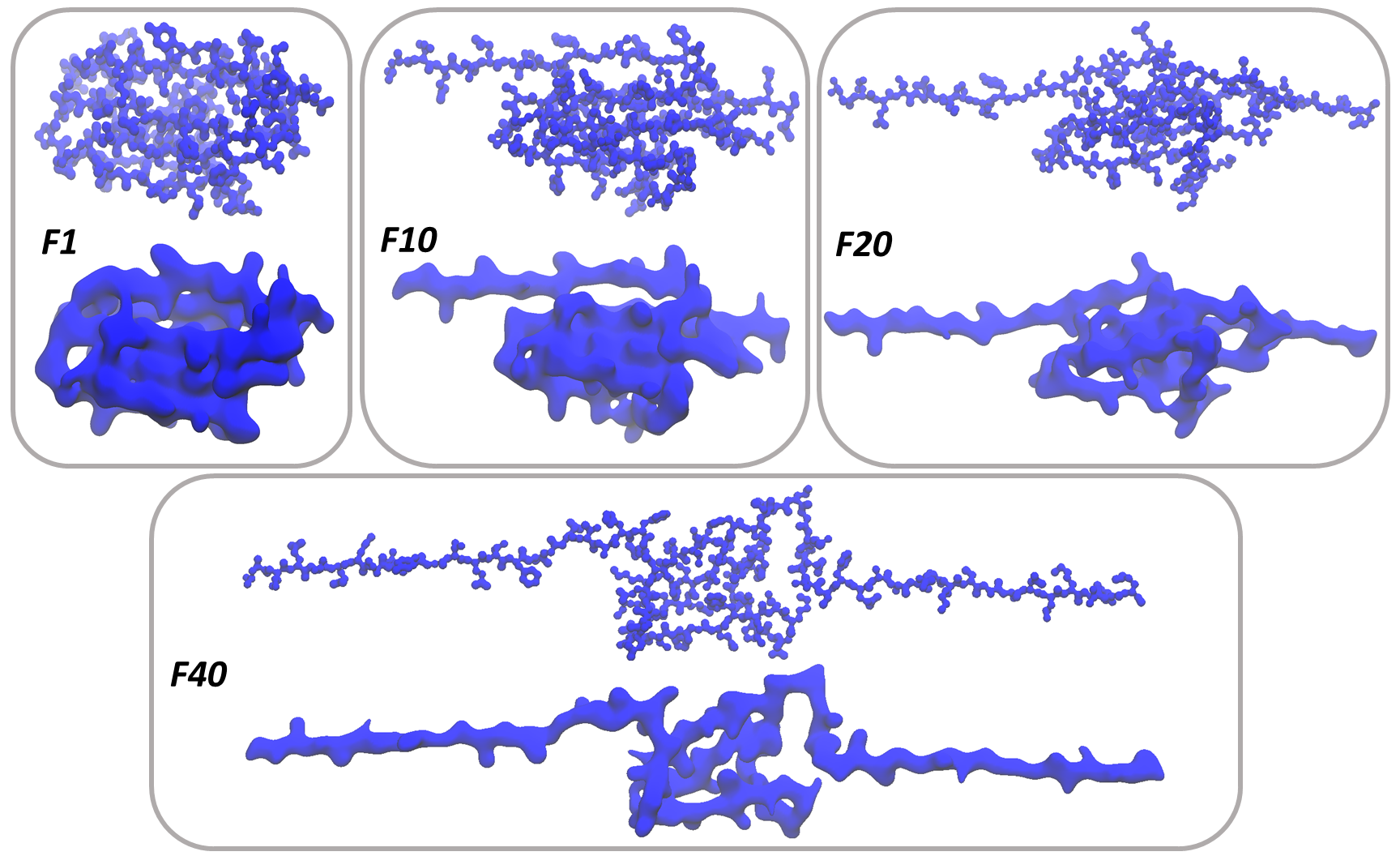}
\end{tabular}
\end{center}
\caption{ The Titin protein structures derived from steered dynamic simulations. There are totally 89 frames extracted from the simulation. I choose only the frames 1, 10, 20 and 40, denoted as F1, F10, F20 and F40, respectively. Two different scale parameters are used, i.e., $\sigma=0.6$ \AA~ and $2.0$\AA,  in the upper and lower figures, respectively. }
\label{fig:titin_structure}
\end{figure}

\begin{figure}
\begin{center}
\begin{tabular}{c}
\includegraphics[width=0.8\textwidth]{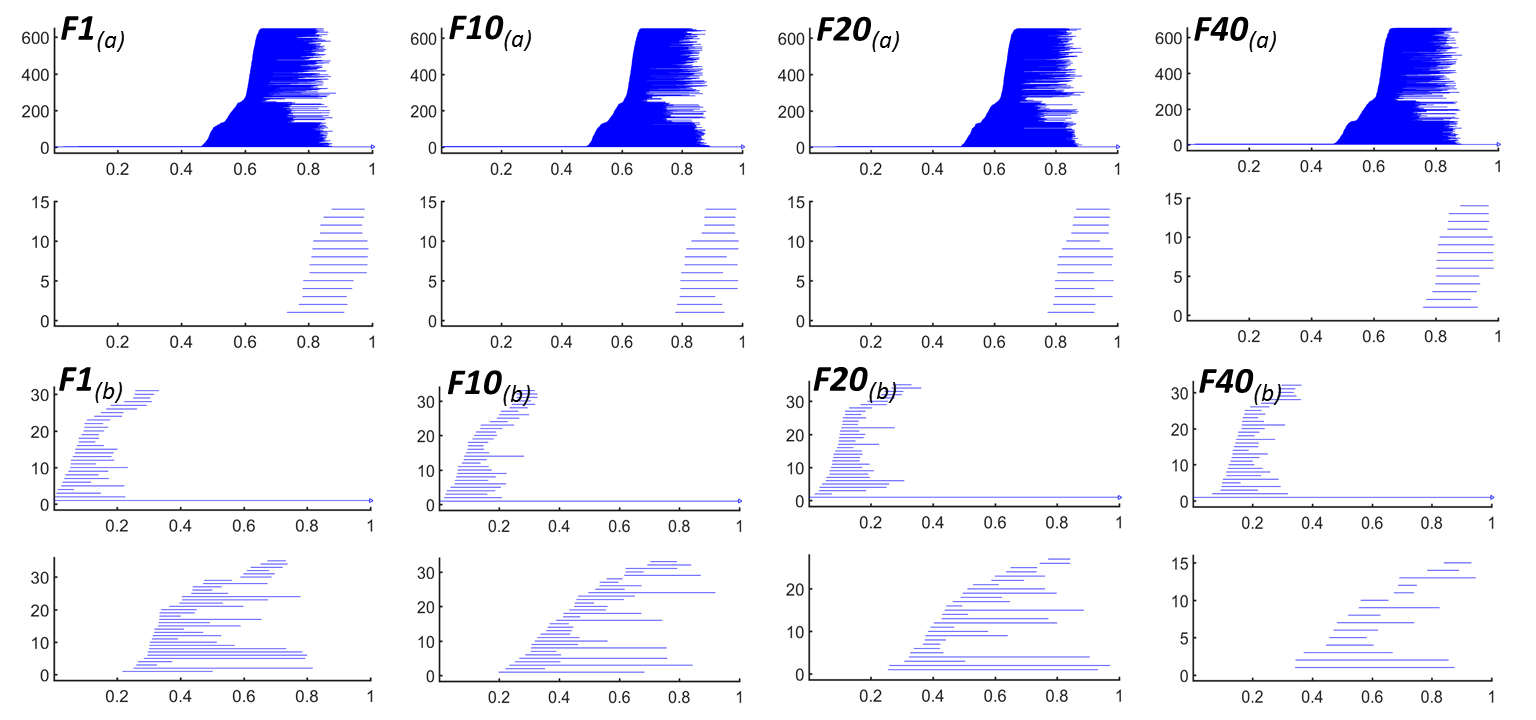}
\end{tabular}
\end{center}
\caption{ Persistent barcodes for the four configurations in Figure \ref{fig:titin_structure}. }
\label{fig:titin_barcode}
\end{figure}

In third case, I consider a classic steered molecular dynamic simulation, i.e., Titin I91. The simulation data are directly taken from the VMD timeline tutorial files from \url{http://www.ks.uiuc.edu/Training/Tutorials/science/timeline/timeline-tutorial-files/}. Four out of eighty-seven frames are picked out for similarity analysis, including frame 1, 10, 20 and 40 (denoted as $F1$ to $F40$). Figure \ref{fig:titin_structure} illustrates the multiscale rigidity functions for these four configurations. I choose the generalized exponential kernels as in Eq.\ref{eq:couple_matrix1}, with parameter $\kappa=2$ and two different resolution values, i.e., $\sigma=0.6$ \AA~ and $2.0$ \AA. The density data is generate with the voxel size of $0.3$ \AA.

Figure \ref{fig:titin_structure} demonstrates the structures of the four frames in two different resolution. The pictures on the upper parts are for resolution value $\sigma=0.6$ \AA~ and low parts for $\sigma=2.0$ \AA. The extension process goes from $F1$ to $F40$ and Titin structure is gradually unfolded accordingly. Since different resolution results in density data in different scales, it enables us to observe variations of structure properties from various scales.

Figure \ref{fig:titin_barcode} illustrates the barcodes for different configurations. Similar to Figure \ref{fig:titin_structure}, the upper figures indicated as ({\bf a}) are for resolution value $\sigma=0.6$ \AA~ and low figures indicated as ({\bf b}) are for $\sigma=2.0$ \AA. It can be seen that, barcodes are much more consistent among four configurations in lower resolution. The barcode lengths, total numbers, and basic patterns show great similarity. In contrast, when resolution enlarges to $\sigma=2.0$ \AA, I begin to observe more variations, particularly in $\beta_1$, its total number keeps decreasing as the structure unfolds.

To have a more quantitative comparison, I calculate the persistent similarity between the first frame and the other three. When $\sigma=0.6$ \AA, I have $P(F1, F10)=(0.940, 0.915)$,  $P(F1, F20)=(0.915, 0.924)$ and $P(F1, F40)=(0.967, 0.975)$.  When $\sigma=2.0$ \AA, I have $P(F1, F10)=(0.832, 0.822)$,  $P(F1, F20)=(0.828, 0.695)$ and $P(F1, F40)=(0.573, 0.413)$. Previously I found that for higher resolution ($\sigma=0.6$ \AA), $\beta_0$ shows atomic information, longer bars represent atoms with larger atomic number and shorter bars are the inverse\cite{Xia2015:multiresolution,KLXia:2015d}. For $\beta_1$, it captures the pentagon and hexagon rings in the aromatic residues. Since atomic information and residues are unchanged during the unfolding process, one will expect a higher persistent similarity in higher resolution structure density data. However, when resolution becomes lower, the fine detailed structure properties are gradually faded away, leaving only the global types of properties. In this case, variations in the folding process should be best represented by the global properties with a lower resolution. My persistent similarity results again confirm my analysis. For higher resolution situation, persistent similarities are very high with values larger than 0.91 for all cases. In lower resolution cases, I find a much smaller similarities, particularly for $\beta_1$. And the values are systematically decreasing, indicating a gradual derivation from the original structure.

It is worth mentioning that even though I have relatively large persistent similarities for higher resolution cases, their persistent similarity is not exactly equal to 1.0 (or just little variation from 1.0). Theoretically, I should have identical $\beta_0$ and $\beta_1$ persistent similarity values as the $\beta_0$ persistent similarity in Case 2. However, due to computational constraints, I only afford to use voxel size of $0.3$ \AA. In this way, the highest rigidity values for frame 1, 10 ,20 and 40 are 15.01, 15.53, 15.03 and 15.30, respectively. This variations induce inconsistence in the normalized rigidity function and further into the barcode results. The volex size will also have some influence on the results from lower resolution situations.

\subsection{Case 4: Fullerene $C_{44}$ isomers}

\begin{figure}
\begin{center}
\begin{tabular}{c}
\includegraphics[width=0.90\textwidth]{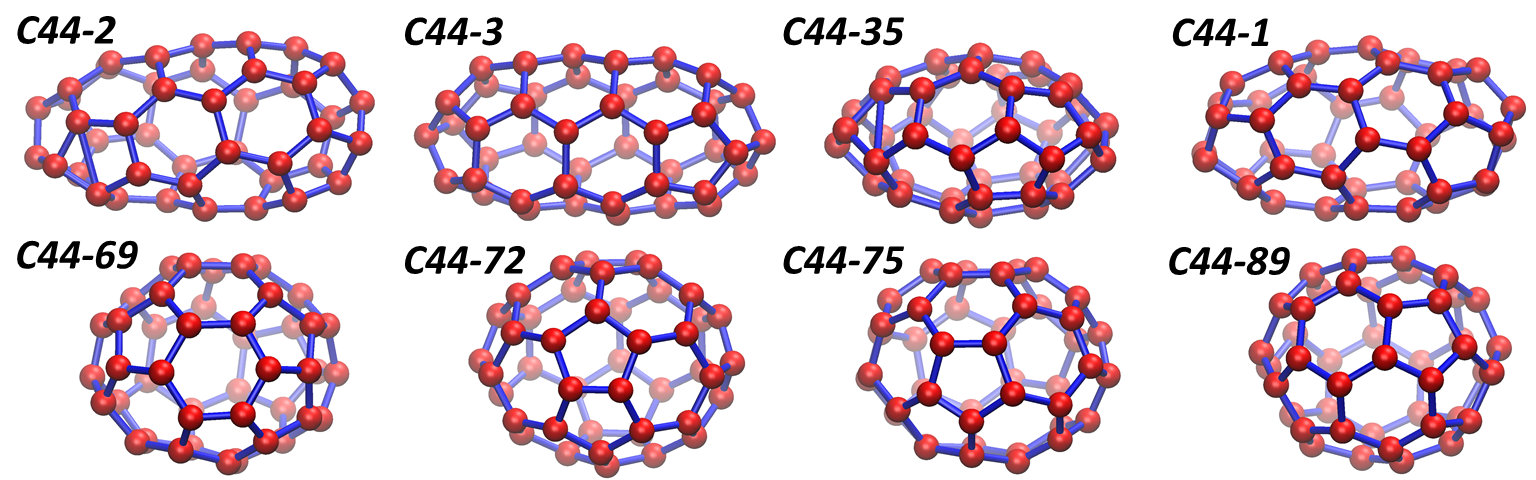}
\end{tabular}
\end{center}
\caption{ The illustration of eight different fullerene $C_{44}$ isomer structures. Fullerene $C_{44}$ has totally 89 isomers. These isomers have different total curvature energies. I have demonstrated four isomer structures with the largest total curvature energies in the upper figures. From the large energies to small ones, their indexes are 2, 3, 35 and 1, respectively. I have illustrated four isomer structures with the smallest total curvature energies in the lower figures. Again from the large energies to small ones, their indexes are 69, 72, 75 and 89, respectively.}
\label{fig:c44_structures}
\end{figure}

%

\begin{figure}
\begin{center}
\begin{tabular}{c}
\includegraphics[width=0.5\textwidth]{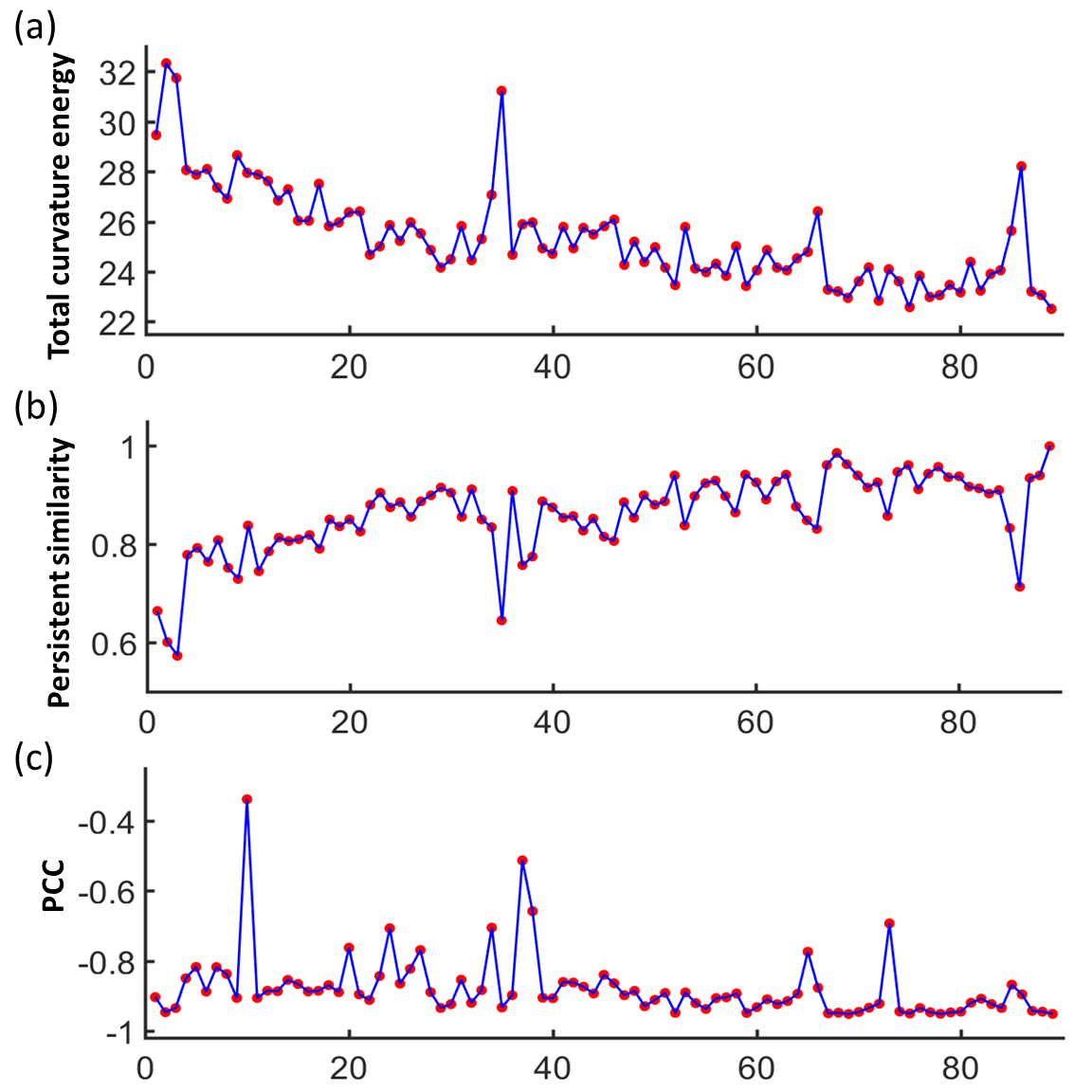}
\end{tabular}
\end{center}
\caption{ The comparison between total curvature energies and persistent similarities of 89 fullerene $C_{44}$ isomers. ({\bf a}) is the total curvature energies of 89 fullerene $C_{44}$ isomers. ({\bf b}) is the persistent similarities between all isomers with the isomer 89. ({\bf c}) The Pearson correlation coefficient (PCC) of persistent similarities and curvature energy differences. For each isomer, I calculate its persistent similarities with all isomers including itself, then I compare these similarity values with the corresponding curvature energy differences (absolute value) to get PCC values.   }
\label{fig:c44_pcc}
\end{figure}

In the last cases, I consider the fullerene $C_{44}$ isomers and its total curvature energies. The fullerene $C_{44}$ isomers structure and energy data can be downloaded from \url{http://www.nanotube.msu.edu/fullerene/fullerene.php?C=44}. To have a general idea how their structures look like, eight isomer structures are specially chosen and illustrated in Figure \ref{fig:c44_structures}. Among them, four isomers with the largest total curvature energies are shown in the upper figures. From the large energies to small ones, their indexes are 2, 3, 35 and 1, respectively. Four isomer structures with the smallest total curvature energies are depicted in the lower figures. Again from the large energies to small ones, their indexes are 69, 72, 75 and 89, respectively.

It is found that the isomer total curvature energy is highly related to the regularity of isomer cage structure. The longest $\beta_2$ barcode, representing the cage size, has been found to be linearly related to these energies\cite{KLXia:2015a}. In this case, I further explore the relation of structure similarities and total curvature energy differences. To remove the irrelevant topological properties, in my PBFs, only the weight for the longest $\beta_2$ barcode is chosen as 1.0 with all the others set as 0.0. And I only consider the $\beta_2$ PBFs. First, I compare the similarity between isomer $C_{44}$-${89}$ and all isomers, and its relation with the total curvature energies. Figure \ref{fig:c44_pcc} ({\bf a}) and ({\bf b}) illustrate the results of energies and similarities, respectively. It can be seen that there is inverse relation between them. Actually, the Pearson correlation coefficient (PCC) between them are -0.952, which is better that distance filtration \cite{KLXia:2015a} and density filtration results \cite{BaoWang:2016a}, and as good as correlation matrix results\cite{KLXia:2015a}. 

Further, I change the reference isomer form $C_{44}$-${89}$ to others and recalculate the PCC between persistent similarity values and curvature energy differences. To avoid confusion, the energy differences are taken as the absolute difference value and are always positive. The PCCs are illustrates in Figure \ref{fig:c44_pcc} ({\bf c}). It can be seen that most of the PCCs are larger than 0.80. More interesting, using the isomers with more extreme curvature energies as the reference gives higher PCCs. In contrast, if the reference isomer is chosen from the ones with intermediate curvature energies, there will be a small PCC value. This is reasonable because similarity measures the ``absolute" different.
Therefore, one should always go to the extreme cases for similarity evaluation so that it can bring out more intrinsic differences. It should be noticed that a Wasserstein metric based similarities may have the potential to avoid this problem. However, this falls out of the scope of the current paper and will be further explored later.

\section{Conclusion remarks}
In this paper, I introduce a persistent similarity model for structure comparison. Based on the persistent homology analysis, the persistent similarity can deliver a quantitative comparison of the intrinsic topological properties. In this model, a persistent Betti function is proposed to represent the barcodes from the persistent homology analysis into a series of one-dimensional functions. The similarity is defined as the ratio of the sizes of intersect areas and union areas from these functions. Further, in order to avoid the ambiguity of comparing structures with no significant topological properties, a pseudo-barcode is introduced. Moreover, multiscale persistent similarity is also introduced to facilitate the comparison of structure properties in different scales. The persistent similarity method is validated with several test examples. It is found that the persistent similarity can be used to describe the intrinsic similarities and differences between the structures very well.

The proposed persistent similarity has several unique properties. Firstly, with the representation of structures in PBFs, the comparison between various structure can be done very efficiently.
In my persistent similarity, any complicated biomolecular structure can be reduced to several simple 1D PBFs for comparison. Their similarity is then defined as the quotient between sizes of intersect areas and union areas below two correspondingly PBFs. Secondly, the multiscale persistent similarity enables an objective-oriented comparison. Through the multiscale rigidity function, a multiscale biomolecular representation is achieved and naturally induces a multiscale persistent similarity. Since my multiscale representation can be adjusted to any resolutions, the associated persistent similarity can be used to compare structures in any particular scale of interest. Thirdly, a pseudo-barcode is introduced to deliver a more precise comparison in the special situation when structures have no significant topological properties.

In future, I will explore the application of persistent similarity in protein structure classification \cite{Rogen2003:automatic} and compare with existing methods to demonstrate its full potential. Further, I will consider Wasserstein metric as a new similarity measurement.

\section*{Acknowledgments}

This work was supported by NTU SUG-M4081842.110 and MOE AcRF Tier 1 M401110000.

\vspace{0.6cm}


\end{document}